\documentclass[twocolumn, times, tighten, appendixfloats]{aastex631}

\usepackage{natbib}
\usepackage{enumitem}
\usepackage{amsmath}
\usepackage{soul}
\usepackage{longtable}

\usepackage{graphicx}		
\usepackage{latexsym}		
\usepackage{bm}			
\usepackage[english]{babel}
\usepackage{color}		
\newcommand{\ie}	{i.e.,}%
\newcommand{\etal}	{\hbox{et~al.}}%

\newcommand{\DEL}[1]{}
\newlength{\txw}
\newlength{\txh}

{\relax}%

\begin{document}

\title{JWST's PEARLS: Bright 1.5--2.0~$\bm{\mu}$m Dropouts in the Spitzer/IRAC Dark Field}

\author[0000-0001-7592-7714]{Haojing Yan} 
\affiliation{Department of Physics and Astronomy, University of Missouri, Columbia, MO 65211, USA}

\author[0000-0003-3329-1337]{Seth H.~Cohen} 
\affiliation{School of Earth \& Space Exploration, Arizona State University, Tempe, AZ 85287-1404, USA}
\author[0000-0001-8156-6281]{Rogier A.~Windhorst} 

\affiliation{School of Earth \& Space Exploration, Arizona State University, Tempe, AZ 85287-1404, USA}
\affiliation{Department of Physics, Arizona State University, Tempe, AZ 85287-1504, USA}

\author[0000-0003-1268-5230]{Rolf A.~Jansen} 
\affiliation{School of Earth \& Space Exploration, Arizona State University, Tempe, AZ 85287-1404, USA}

\author[0000-0003-3270-6844]{Zhiyuan Ma}
\affiliation{Department of Astronomy, University of Massachusetts, Amherst, MA 01003, USA}

\author[0000-0002-0005-2631]{John F. Beacom} 
\affiliation{Center for Cosmology and AstroParticle Physics (CCAPP), The Ohio State University, Columbus, OH 43210, USA}
\affiliation{Department of Physics, The Ohio State University, Columbus, OH 43210, USA}
\affiliation{Department of Astronomy, The Ohio State University, Columbus, OH 43210, USA}

\author[0000-0003-4952-3008]{Chenxiaoji Ling}
\affiliation{Department of Physics and Astronomy, University of Missouri, Columbia, MO 65211, USA}

\author[0000-0003-0202-0534]{Cheng Cheng}
\affiliation{Chinese Academy of Sciences South America Center for Astronomy, National Astronomical Observatories, CAS, Beijing 100101, China}

\author[0000-0001-6511-8745]{Jia-Sheng Huang}
\affiliation{Chinese Academy of Sciences South America Center for Astronomy, National Astronomical Observatories, CAS, Beijing 100101, China}

\author[0000-0001-9440-8872]{Norman A.~Grogin} 
\affiliation{Space Telescope Science Institute, 3700 San Martin Drive, Baltimore, MD 21218, USA}

\author[0000-0002-9895-5758]{S. P. Willner}
\affiliation{Center for Astrophysics \textbar\ Harvard  \& Smithsonian, 60 Garden St., Cambridge, MA 02138, USA}

\author[0000-0001-7095-7543]{Min Yun}
\affiliation{Department of Astronomy, University of Massachusetts, Amherst, MA 01003, USA}

\author[0000-0001-8751-3463]{Heidi B.~Hammel} 
\affiliation{Association of Universities for Research in Astronomy, Washington, DC 20005, USA}

\author[0000-0001-7694-4129]{Stefanie N.~Milam} 
\affiliation{NASA Goddard Space Flight Center, Greenbelt, MD\,20771, USA}

\author[0000-0003-1949-7638]{Christopher J. Conselice} 
\affiliation{Jodrell Bank Centre for Astrophysics, Alan Turing Building, 
University of Manchester, Oxford Road, Manchester M13 9PL, UK}

\author[0000-0001-9491-7327]{Simon P. Driver} 
\affiliation{International Centre for Radio Astronomy Research (ICRAR) and the
International Space Centre (ISC), The University of Western Australia, M468,
35 Stirling Highway, Crawley, WA 6009, Australia}

\author[0000-0003-1625-8009]{Brenda Frye} 
\affiliation{Steward Observatory, University of Arizona, 933 N Cherry Ave,
Tucson, AZ, 85721-0009, USA}

\author[0000-0001-6434-7845]{Madeline A. Marshall} 
\affiliation{National Research Council of Canada, Herzberg Astronomy \&
Astrophysics Research Centre, 5071 West Saanich Road, Victoria, BC V9E 2E7, 
Canada}
\affiliation{ARC Centre of Excellence for All Sky Astrophysics in 3 Dimensions
(ASTRO 3D), Australia}

\author[0000-0002-6610-2048]{Anton Koekemoer}
\affiliation{Space Telescope Science Institute, 3700 San Martin Drive,
Baltimore, MD 21218, USA}

\author[0000-0001-9262-9997]{Christopher N. A. Willmer}  
\affiliation{Steward Observatory, University of Arizona, 933 N Cherry Ave, Tucson, AZ, 85721-0009}

\author[0000-0003-0429-3579]{Aaron Robotham}
\affiliation{International Centre for Radio Astronomy Research (ICRAR) and the
International Space Centre (ISC), The University of Western Australia, M468,
35 Stirling Highway, Crawley, WA 6009, Australia}

\author[0000-0002-9816-1931]{Jordan C. J. D'Silva} 
\affiliation{International Centre for Radio Astronomy Research (ICRAR) and the
International Space Centre (ISC), The University of Western Australia, M468,
35 Stirling Highway, Crawley, WA 6009, Australia}

\author[0000-0002-7265-7920]{Jake Summers}
\affiliation{School of Earth \& Space Exploration, Arizona State University,
Tempe, AZ 85287-1404, USA}

\author[0000-0003-4220-2404]{Jeremy Lim} 
\affiliation{Department of Physics, The University of Hong Kong, Pokfulam Road, 
Hong Kong}

\author[0000-0001-5429-5762]{Kevin Harrington}
\affiliation{European Southern Observatory, Alonso de C{\'o}rdova 3107, Vitacura, Casilla 19001, Santiago de Chile, Chile}

\author[0000-0002-8919-079X]{Leonardo Ferreira}
\affiliation{University of Nottingham, School of Physics \& Astronomy, Nottingham, NG7 2RD, UK}

\author[0000-0001-9065-3926]{Jose Maria Diego}
\affiliation{Instituto de Física de Cantabria (CSIC-UC), Avda. Los Castros s/n, 39005, Santander, Spain}

\author[0000-0003-3382-5941]{Nor Pirzkal} 
\affiliation{Space Telescope Science Institute, 3700 San Martin Drive, Baltimore, MD 21218, USA}

\author[0000-0003-3903-6935]{Stephen M. Wilkins} 
\affiliation{Astronomy Centre, Department of Physics and Astronomy, University of Sussex, Brighton, BN1 9QH, UK}

\author[0000-0001-7092-9374]{Lifan Wang}
\affiliation{Department of Physics and Astronomy, Texas A\& M University, Mitchell Physics Building (MPHY) 4242 TAMU, College Station, TX 77843-4242}

\author[0000-0001-6145-5090]{Nimish P. Hathi} 
\affiliation{Space Telescope Science Institute, 3700 San Martin Drive,
Baltimore, MD 21218, USA}

\author[0000-0002-0350-4488]{Adi Zitrin} 
\affiliation{Physics Department, Ben-Gurion University of the Negev, P.O. Box 653, Beer-Sheva 8410501, Israel}

\author[0000-0003-0883-2226]{Rachana A. Bhatawdekar} 
\affiliation{European Space Agency, ESA/ESTEC, Keplerlaan 1, 2201 AZ Noordwijk, NL}

\author[0000-0003-4875-6272]{Nathan J. Adams} 
\affiliation{Jodrell Bank Centre for Astrophysics, Alan Turing Building, 
University of Manchester, Oxford Road, Manchester M13 9PL, UK}

\author[0000-0001-6278-032X]{Lukas J. Furtak} 
\affiliation{Physics Department, Ben-Gurion University of the Negev, P.O. Box 653, Beer-Sheva 8410501, Israel}

\author[0000-0002-2203-7889]{Peter Maksym} 
\affiliation{Center for Astrophysics \textbar\ Harvard  \& Smithsonian, 60 Garden St., Cambridge, MA 02138, USA}

\author[0000-0001-7016-5520]{Michael J. Rutkowski} 
\affiliation{Minnesota State University-Mankato, Trafton North Science Center, 
Mankato, MN, 56001, USA}

\author[0000-0002-0670-0708]{Giovanni G. Fazio}
\affiliation{Center for Astrophysics \textbar\ Harvard  \& Smithsonian, 60 Garden St., Cambridge, MA 02138, USA}


\correspondingauthor{Haojing Yan}
\email{yanha@missouri.edu}


\setwatermarkfontsize{1in}

\shortauthors{Yan \etal}
\shorttitle{Dropouts in the IRAC Dark Field}

\begin{abstract}
   
   Using the first epoch of four-band NIRCam observations obtained by the James
Webb Space Telescope (JWST) Prime Extragalactic Areas for Reionization and 
Lensing Science Program in the Spitzer IRAC Dark Field, we search for F150W and
F200W dropouts. In 14.2~arcmin$^2$, we have found eight F150W dropouts and 
eight F200W dropouts, all brighter than 27.5~mag (the brightest being 
$\sim$24~mag) in the band to the red side of the break. As they are detected in
multiple bands, these must be real objects. Their nature, however, is unclear,
and characterizing their properties is important for realizing the full
potential of JWST\null. If the observed color decrements are due to the Lyman
break, these objects should be at $z\gtrsim 11.7$ and $z\gtrsim 15.4$, 
respectively. The color diagnostics show that at least four F150W dropouts are 
far away from the usual contaminators encountered in dropout searches (red 
galaxies at much lower redshifts or brown dwarf stars). While the diagnostics
of the F200W dropouts are less certain due to the limited number of passbands,
at least one of them is likely not a known type of contaminant, and the rest
are consistent with either high-redshift galaxies with evolved stellar
populations or old galaxies at $z\approx 3$ to 8. If a significant fraction of
our dropouts are indeed at $z\gtrsim 12$, we have to face the severe problem of
explaining their high luminosities and number densities. Spectroscopic 
identifications of such objects are urgently needed.

\end{abstract}

\keywords{dark ages, reionization, first stars --- galaxies: evolution
--- galaxies: high-redshift} 


\section{Introduction}

   The advent of the James Webb Space Telescope (JWST) has pushed our
redshift frontier to $z>11$ and even possibly to $z\approx 20$. This is enabled
by its NIRCam instrument, which offers wavelength coverage far beyond the 
1.7~$\mu$m cutoff of the Hubble Space Telescope (HST) and thus allows
objects at $z>11$ to be selected. The first batch of deep NIRCam data, released
on 2022 July 14, immediately spurred many independent groups to search for
objects at the highest possible redshifts. These data were obtained in three
different fields of similar coverage, namely, the JWST Early Release 
Observations \citep[ERO;][]{Pontoppidan2022} in the SMACS J0723$-$73 cluster
field (hereafter ``SMACS0723''), the GLASS JWST
Early Release Science Program \citep[][]{Treu2022} in the 
flanking field of Abell 2744 (hereafter ``GLASS''), and the Cosmic Evolution 
Early Release Science Survey (hereafter ``CEERS''; Finkelstein et al.\ in prep.)\ 
in the Extended Groth Strip. Within a month after these data were released, over
a hundred candidate $z>11$ galaxies have been reported 
\citep[][]{Naidu2022a, Castellano2022, Adams2022, Yan2022b, Atek2022, 
Donnan2022, Finkelstein2022b, Rodighiero2022, Harikane2022} using either the 
dropout method that identifies Lyman-break or the photometric redshift
($z_{\rm ph}$) method. The largest sample, from \citet[][]{Yan2022b}, 
contains 87 candidates up to $z\approx 20$ in SMACS0723. In stark contrast, 
years of searches based on the HST near-infrared (NIR) data in multiple
fields resulted in only one $z\approx 11$ galaxy, which could be at 
$z=11.09$ \citep[][]{Oesch2016} or 10.957 \citep[][]{Jiang2021}.

   While the early JWST searches are a giant leap forward, severe 
problems have also surfaced. Most studies of galaxy formation in the early 
universe did not predict such a large number of $z>11$ galaxies to be found.
Furthermore, most of the aforementioned $z>11$ candidate samples contain some
very bright objects that are difficult to reconcile with our current 
understanding of early galaxy formation processes. For example, the 
$z\approx 11$--20 candidates reported by \cite{Yan2022b} in SMACS0723 include
five objects that are brighter than 26.5~mag in F356W\null. Two of the candidates
reported by \cite{Atek2022} in the same field have F200W magnitudes of 25.22
and 26.35 at $z_{\rm ph}=11.22$ and 15.70, respectively. Among the candidates
reported by \cite{Castellano2022}, there is also a very bright object at
$z_{\rm ph}=12.3$ in GLASS with F444W magnitude of 25.88. In CEERS, 
\cite{Donnan2022} found a candidate at $z_{\rm ph}=16.74$ that has an F200W 
magnitude of 26.46. All such objects, if at the high redshifts suggested, would 
correspond to $M_{\rm UV}\lesssim -21.5$ to $-24$, a regime where no previous 
studies had suggested finding any galaxies over such a small area as a few NIRCam pointings. 
Gravitational lensing cannot solve the problem, as most of these objects have
no evidence suggesting significant lensing. Even near the lensing cluster in
the SMACS0723 field, only a couple of the $z>11$ candidates could be magnified
by a factor of $\sim 3$ \citep[][]{Yan2022b}. On the other hand, it is not
impossible that most of these very bright candidates are due to some novel
kinds of contaminators; in this case, such objects are worth further
investigation
in their own right and
so that JWST high-redshift studies can be put on a
solid footing.

   Given the tension already created by these surprising initial results, it is
important to verify whether such bright $z>11$ {\it candidates}\ are also seen
in other fields. To this end, we report our initial search for dropouts
in one of the ``blank'' fields of the Prime Extragalactic Areas for 
Reionization and Lensing Science program \citep[PEARLS;][]{Windhorst2022},
which is a JWST Interdisciplinary Scientists Guaranteed Time Observation
Program (PI. Windhorst; PID 1176 \& 2738). We describe our NIRCam data and
the source extraction in Section~2. We focus on the dropouts from two NIRCam
bands (F150W and F200W at $\sim$1.5 and $\sim$2.0~$\mu$m, respectively), which
are presented in Section~3. We conclude with a discussion in Section~4. All 
magnitudes are in the AB system, and we adopt a flat $\Lambda$CDM cosmology 
with parameters 
$H_0=71$~km~s$^{-1}$~Mpc$^{-1}$, $\Omega_M=0.27$, and $\Omega_\Lambda=0.73$.


\section{Observations and Data Reduction}

    The PEARLS NIRCam data used in this study are in the central portion of
the Spitzer IRAC Dark Field \citep[IDF;][]{Krick2009, Yan2018}, 
dubbed the ``JWIDF'' (Yan \etal\ 2022b, in preparation). This field is in the
JWST continuous viewing zone and has deep prior observations from the 
IRAC camera on the Spitzer Space Telescope for 16.2 years. This 
PEARLS field was chosen primarily for the IR time-domain science and has three
planned epochs of 4-band NIRCam observations in F150W, F200W, F356W, and F444W. 

    The current work is based on the first epoch of observations, which were
executed on 2022 July 8~UT\null. NIRCam operates in the ``short wavelength''
(SW) and ``long wavelength'' (LW) channels simultaneously, and we paired 
observations in the F150W band with F444W and the F200W band with F356W. To 
cover the gaps between detectors, we used \texttt{FULLBOX} dithers with the
\texttt{6TIGHT} pattern, which results in a $\sim$5\farcm9$\times$2\farcm4
rectangle area covered by six dithered exposures. The dithered positions are
determined by the \texttt{STANDARD} subpixel dither to optimally sample the
point spread functions (PSFs). For each exposure, the \texttt{SHALLOW4} readout
pattern was adopted with ``up-the-ramp'' fitting to determine the count rate.
We used one integration per exposure with 10 groups per integration, giving a
uniform exposure time of 3157 seconds in each band. The native NIRCam pixel 
scales are 0\farcs031~pix$^{-1}$ for F150W and F200W (both in SW) and 
0\farcs063~pix$^{-1}$ for F356W and F444W (both in LW).

    The data were retrieved from the Mikulski Archive for Space Telescopes
(MAST)\null. Reduction started from the so-called Stage~1 ``uncal'' products,
which are the single exposures from the standard JWST data reduction pipeline 
after Level~1b processing. The JWST data reduction pipeline has been quickly
evolving, and we used the version 1.6.1dev3+gad99335d in the context of 
\texttt{jwst\_0944.pmap},\footnote{Explanation of JWST calibration versions is at 
\url{https://jwst-docs.stsci.edu/jwst-calibration-pipeline-caveats}.} 
which takes the latest NIRCam flux calibrations (as of 2022-08-20) into account.
A few changes and augmentations were made to the pipeline to improve the 
reduction quality; most importantly, these included enabling the use of an 
external reference catalog for image alignment and implementing a better 
background estimate for the final stacking. Removal of the so-called ``1/f''
patterns in the SW bands was also integrated in the process. The single
exposures in each band were stacked and were projected onto the same 
astrometric grid with a pixel scale of 0\farcs06 (hereafter the ``60mas'' 
version). This choice of scale sacrifices some angular resolution in the SW 
bands in favor of better detection of faint sources. The mosaics are in surface
brightness units of MJy~sr$^{-1}$. For the 0\farcs06 pixel scale, this 
translates to a magnitude zeropoint of 26.581. In addition, we also created 
another version of mosaics at a pixel scale of 0\farcs03 (hereafter the
``30mas'' version), which were used to study the sizes of the selected 
candidates.

\begin{figure*}[t]
\epsscale{1.2}
\plotone{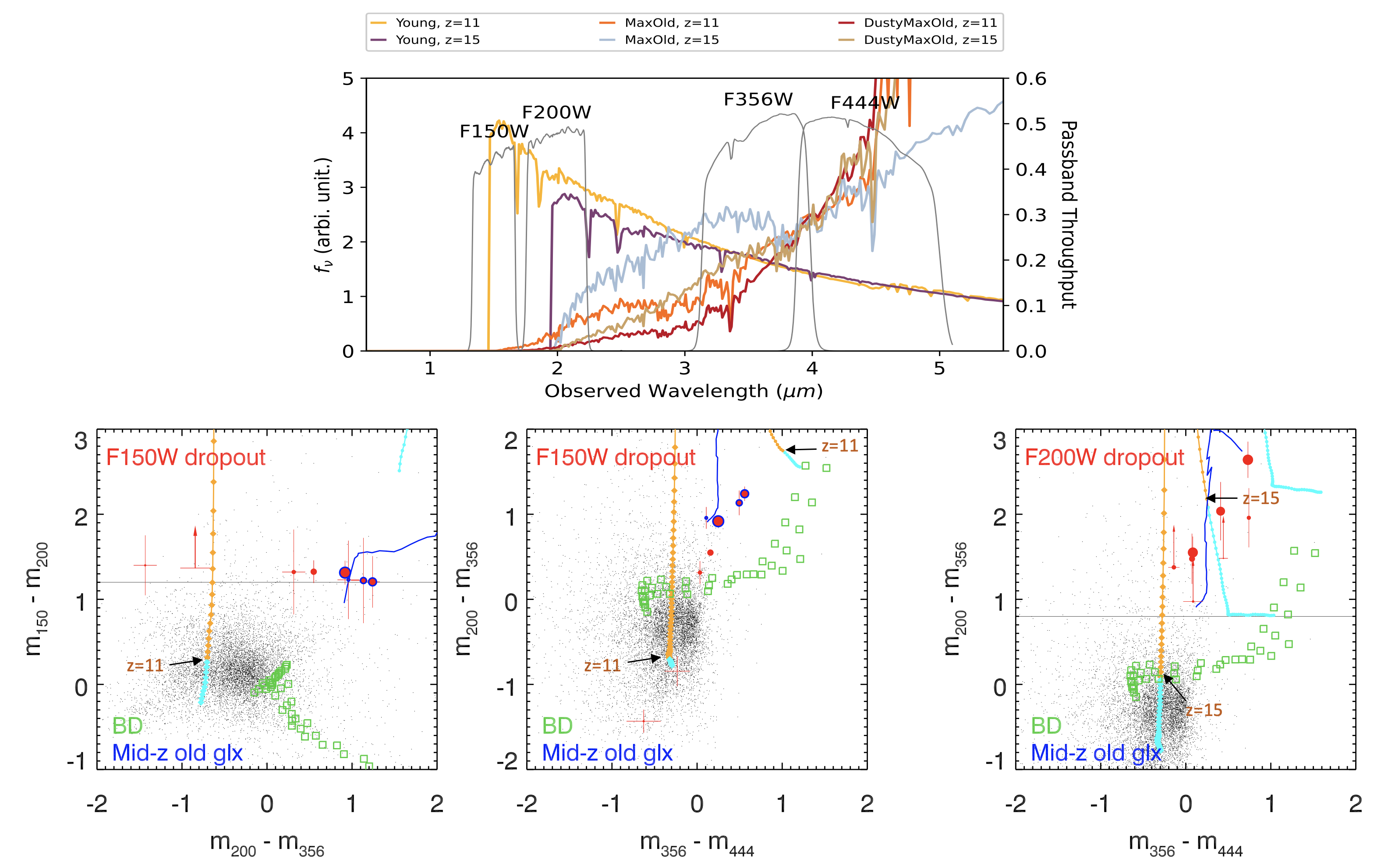}
\caption{(top) Passbands of the JWIDF observations with superposed \citet{Bruzual2003}
templates (all with Chabrier 2003 initial mass functions and solar 
metallicities). Gold and dark brown show 10~Myr templates at $z=11$ and 
$z=15$, respectively. Red and cyan show maximally old (i.e., as old as the age
of the Universe) galaxies at the same
redshifts, and dark red and orange the same maximally old templates with 
$A_V=2.0$~mag (\citealt{Calzetti2001} extinction law). 
(bottom) Diagnostic color--color diagrams for dropout selections. The left 
and middle panels are for F150W dropouts, and the right panel shows the F200W 
dropouts.  The filled red circles with error bars or limits represent the 
selected dropouts, with symbol sizes corresponding to $m_{200}$ (left and 
middle) or $m_{356}$ (right). Black dots show field objects. Colored lines with
dots show the color tracks from $z=10$ to 30 of the three model templates in 
the top panel. The nearly vertical track is the young template, and cyan and
orange indicate respectively  $z<11$, $z>11$ in the left and middle panels and
$z<15$, $z>15$ in the right panel. The tracks of the old templates are shown 
similarly, but the tracks are mostly outside the plotted color ranges.
The blue curve is the color track of mid-$z$ maximally-old galaxies 
from $z=2.6$ to 8.0. This track is shown in its entirety in the right panel 
but is only partially inside the left and middle panels. Four dropouts in the 
left and middle panels marked with blue circles are near these tracks and are
likely mid-$z$ contaminants. The green squares show the colors of
brown dwarfs (labeled by ``BD'') based on \citet[][]{Burrows2006}. 
}
\label{fig:2color}
\end{figure*}

We carried out source extraction and photometry using SExtractor 
\citep{Bertin1996} in dual-image mode. Following \citet[][]{Yan2022b},
we used the F356W image for detection and adopted \texttt{MAG\_ISO} magnitudes
for color measurements. The F356W image is the deepest, and its point spread 
function is comparable to that in F444W but is almost twice as large as those
in the two SW bands. The sources of interest are small enough that the F356W 
\texttt{MAG\_ISO} apertures  include nearly all the source flux while 
minimizing the background noise. Hereafter we denote the magnitudes in 
the four bands as $m_{150}$, $m_{200}$, $m_{356}$, and $m_{444}$, respectively. 
To minimize false detections, we kept only the sources that have 
${\rm S/N}\geq 5.0$ and \texttt{ISOAREA\_IMAGE} $\geq 10$~pixels in F356W.


\section{Dropout Selection}

   The dropout method has been widely accepted as a robust technique of 
selecting Lyman-break galaxy candidates even when only limited bands are 
available. Our motivation was to select $z>11$ candidates, and we followed the
standard procedures. However, we caution that the resulting dropouts should
be treated as nothing more than candidates. Our goal was to verify whether we
could find similar, bright $z>11$ candidates seen in other fields, some of 
which were found using different methods. 

\subsection{Ancillary HST Data}

   In addition to the JWST NIRCam data, we also made use of the archival
HST data taken by the Advanced Camera for Surveys (ACS), which were obtained
in 2006 November 27 UT (PI. Surace; HST PID 10521) in the IDF. These
observations were done in F814W ($\sim$0.806~$\mu$m) at 2-orbit depth
(effective exposure time $\sim$5176 seconds). While they are not as deep as
the NIRcam data, these ACS images are still useful in rejecting the brightest
contaminants. We created a mosaic covering the JWIDF footprint, which was 
registered to the same grid as the 60mas NIRCam mosaics. The nominal 2$\sigma$
depth within 0\farcs2 radius aperture is 28.47~mag.

\subsection{Selection Overview}

   The color criteria for  F150W dropouts and F200W dropouts were chosen 
following the methodology of \citet{Yan2022b}. A flat spectrum (in $f_\nu$) is
typical for Lyman-break galaxies at high redshifts. If such a spectrum is 
truncated at the midpoint of a passband (the ``drop-out band''), the color 
decrement between that band and a redder one is 0.75~mag.  We therefore adopted
a simple color threshold of 0.8~mag, \ie\ $m_{200}-m_{356}\geq 0.8$~mag, to 
select F200W dropouts. A further requirement for F200W dropouts was that the 
source must have ${S/N\leq 2}$ in the ``veto band'' (F150W)\null. When 
selecting F150W dropouts, we do not have a bluer NIRCam band to serve as a 
veto, and therefore we required $m_{150}-m_{200}\geq 1.2$~mag to reduce the 
chance of contamination. This threshold is equivalent to detecting a sharp 
break when it moves $>$2/3 out of the drop-out band. If the break is the
Lyman break, these criteria correspond to $z\gtrsim 11.7$ for F150W dropouts
and $z\gtrsim 15.4$ for F200W dropouts. When calculating the color decrements,
we replaced any ${S/N\leq 2}$ detections in the drop-out band with the 
2-$\sigma$ depths as measured in the \texttt{MAG\_ISO} apertures defined on
the F356W image. A legitimate dropout should also be detected at ${S/N\geq 5}$
in the band to the red side of the break (the ``drop-in'' band) and a null
detection in the ACS F814W image. After the initial selections, we visually
inspected the images of these candidates in all bands to reject contaminators
due to spurious detections around bright objects, image defects, noise spikes 
mistakenly included as sources, etc. Due to photometric errors, some SExtractor
non-detections (below 2-$\sigma$) in the veto band (for F200W dropouts) are in
fact visible; such contaminants were removed in this visual inspection step as
well. 

    Secondary color criteria involving a redder band are often applied in the
dropout selections at lower redshifts (e.g., at $z\approx 6$) to help remove 
possible contaminants such as galaxies with old stellar populations and 
Galactic brown dwarfs. The former have prominent 4000\,\AA\ breaks, and the 
latter have strong molecular absorption bands, both of which could mimic a
Lyman-break signature. As \citet[][]{Yan2022b}, we did not use any such
secondary criteria (but see below for diagnostics). The age of the Universe at
such a high redshift is short enough that activities of short time scales are
not averaged out, and therefore galaxies could have a wide range of colors. 
This is illustrated by the model spectra in the top panel of 
Figure~\ref{fig:2color}, which were generated based on the population synthesis 
models of \citet[][BC03]{Bruzual2003} using the initial mass function (IMF) of
\citet[][]{Chabrier2003} and solar metallicity. Three models are shown. One is
a very young (age of 10~Myr) galaxy with nearly constant star formation 
(hereafter the ``young'' model galaxy), which represents the bluest population
that one can get from BC03. The opposite is a ``maximally old'' template, which
is a single burst (``simple stellar population'' or SSP) whose age is as old 
as the age of the Universe at the redshift under discussion (hereafter the 
``maximally old'' model), e.g., age of 0.5~Gyr (0.3~Gyr) at $z=11$ ($z=15$).
Such a template has the reddest color among the BC03 models. To make it even
redder, we consider a third template, which is a dusty, maximally old 
template with $A_V=2.0$~mag and reddened according to the extinction law of 
\citet[][]{Calzetti2001} (hereafter the ``dusty maximally old'' model).

   We still considered a posteriori the possible impact of the two types of 
aforementioned contaminators. For brown dwarfs, we used a set of model spectra
of \citet[][]{Burrows2006}, which cover L and T brown dwarfs with effective
temperatures ranging from 2300 to 700~K. For old galaxies, we used a series of
BC03 models redshifted to $z=2.6$--8.0 at a stepsize of 0.1.
These models are SSPs with solar metallicity and are ``maximally old,'' i.e.,
their ages are as old as the ages of the Universe at their redshifts. The
population of so-called ``HST-dark'' or ``$H$-band dropout'' galaxies that have
been discussed in the past few years are thought to contain such very old 
components and could have similar near-IR colors as our dropouts 
\citep[e.g.,][and references therein]{Barrufet2022}. We compare our
candidates to all such possible contaminators in the color space.

\begin{figure*}[t]
  \centering
  \includegraphics[scale=0.53]{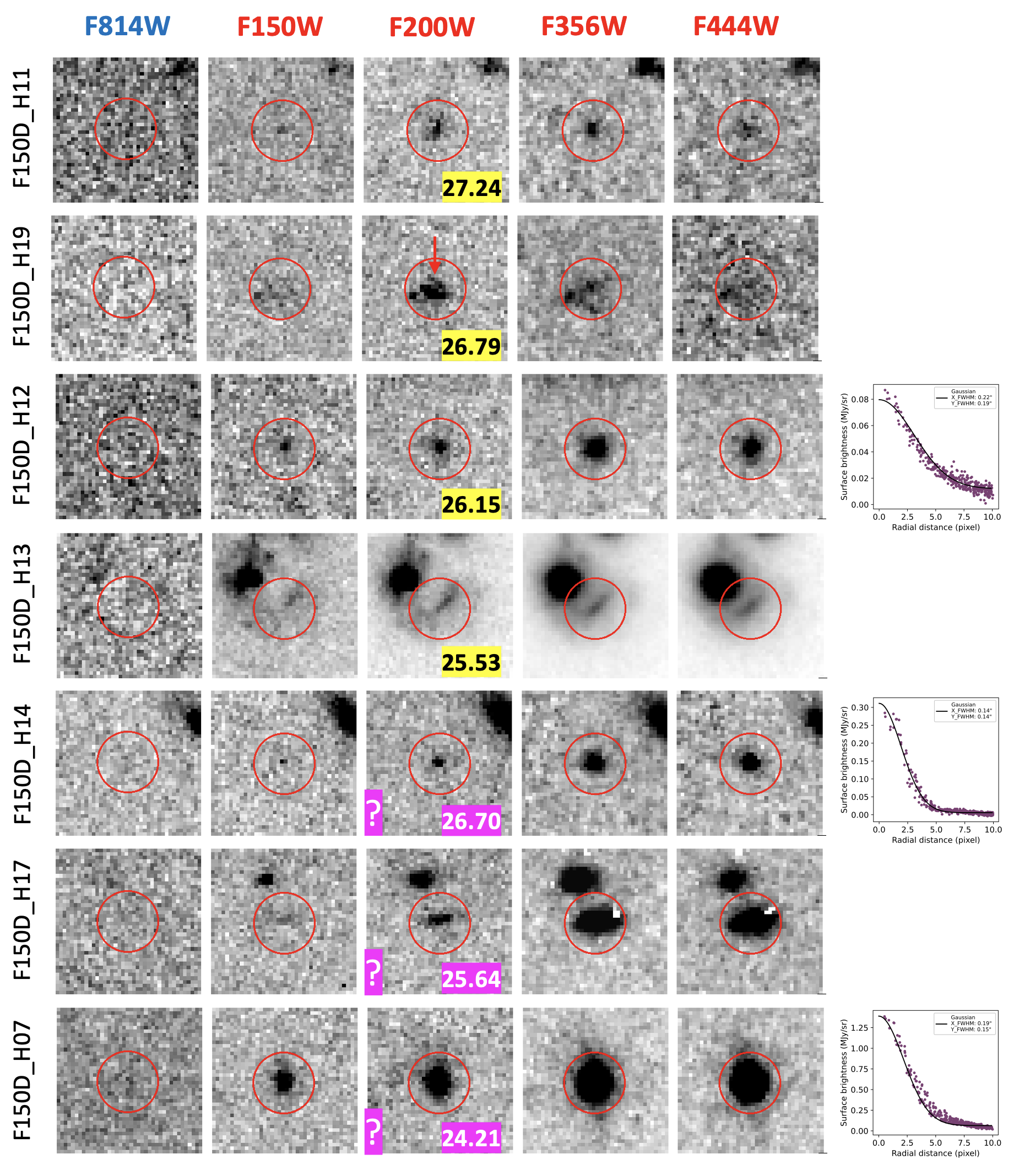}
  \caption{Image stamps of seven of the eight F150W dropouts (with short IDs 
  noted) in the HST ACS F814W and four JWST NIRCam bands (from left to right).
  The eighth is shown in Figure~\ref{fig:special}. The images are 
  2\farcs4$\times$2\farcs4 in size, have 60~mas pixels, and are oriented 
  north-up and east-left. The dropouts are centered on the images and are 
  indicated by red circles (0\farcs5 radius). \texttt{F150D\_H19} has a
  close neighbor that is unrelated, and an arrow is used to indicate the 
  dropout to avoid confusion. The numbers shown on the F200W images are their
  magnitudes in this band ($m_{200}$). The last three objects are among the
  four that are close to the mid-$z$ old galaxy tracks in 
  Figure \ref{fig:2color}, and the question marks are to indicate that they 
  might be contaminants by these diagnostics. \texttt{F150D\_H12}, \texttt{H14},
  and \texttt{H07} are compact, and the last panels show the 2-D Gaussian
  profile fits to their light distributions in F356W. \texttt{H14} is
  consistent with being a point source.
}
  \label{fig:stamps_150d}
\end{figure*}


\subsection {F150W dropouts}

   Our final sample contains eight F150W dropouts. Figure \ref{fig:2color} shows
their colors and compares them to the synthesized colors using the model spectra
in the upper panel as well as those of the two types of possible contaminators
(mid-$z$ old galaxies and brown dwarfs). In the primary selection diagram,
$m_{150}-m_{200}$ versus $m_{200}-m_{356}$, most of the F150W dropouts are far
away from the contamination regions. Brown dwarfs are blue in $m_{150}-m_{200}$
and will never show up as F150W dropouts. These eight dropouts form three 
groups in $m_{200}-m_{356}$ color. Two dropouts (\texttt{F150D\_H11} and
\texttt{F150D\_H19}) are $>$0.5~mag bluer than the young template, which cannot
be caused by contaminators but could be explained by a more top-heavy IMF than
the adopted Chabrier IMF\null. Two dropouts (\texttt{F150D\_H12} and 
\texttt{H13}) have $0.25<m_{200}-m_{356}\leq 0.55$~mag, which could be
explained by high-$z$ models with a range of ages. Finally, four dropouts
(\texttt{F150D\_H14}, \texttt{H17}, \texttt{H07}, and \texttt{E01}) are close
to the track of mid-$z$ old galaxies ($z=2.6$ to 8), suggesting that they could
be contaminants. Their locations in the secondary diagnostic diagram, 
$m_{200}-m_{356}$ versus $m_{356}-m_{444}$, are consistent with this 
interpretation. However, this still cannot rule out with certainty their being
old galaxies at $z>11$.

  Figure~\ref{fig:stamps_150d} shows 5-band stamp images of seven F150W
dropouts. The eighth of them is special and is discussed in 
Section~\ref{s:special}. Their morphologies range from compact to 
diffuse. We fit 2-D Gaussian profiles to the 30mas F356W images of the three 
compact ones (\texttt{F150D\_H12}, \texttt{H14} and \texttt{H02}), and these
best-fit profiles are also shown. \texttt{F150D\_H14} has full-width at
half-maximum of 0\farcs14 in both dimensions and is consistent with being a
point source.
\texttt{F150D\_JWIDF\_H13} appears to be arc-like and could be 
gravitationally lensed by a foreground galaxy that is only 1\farcs2 away, 
although we cannot rule out that it is in fact part of this bright neighbor.


\begin{figure*}[t]
  \centering
  \includegraphics[scale=0.32]{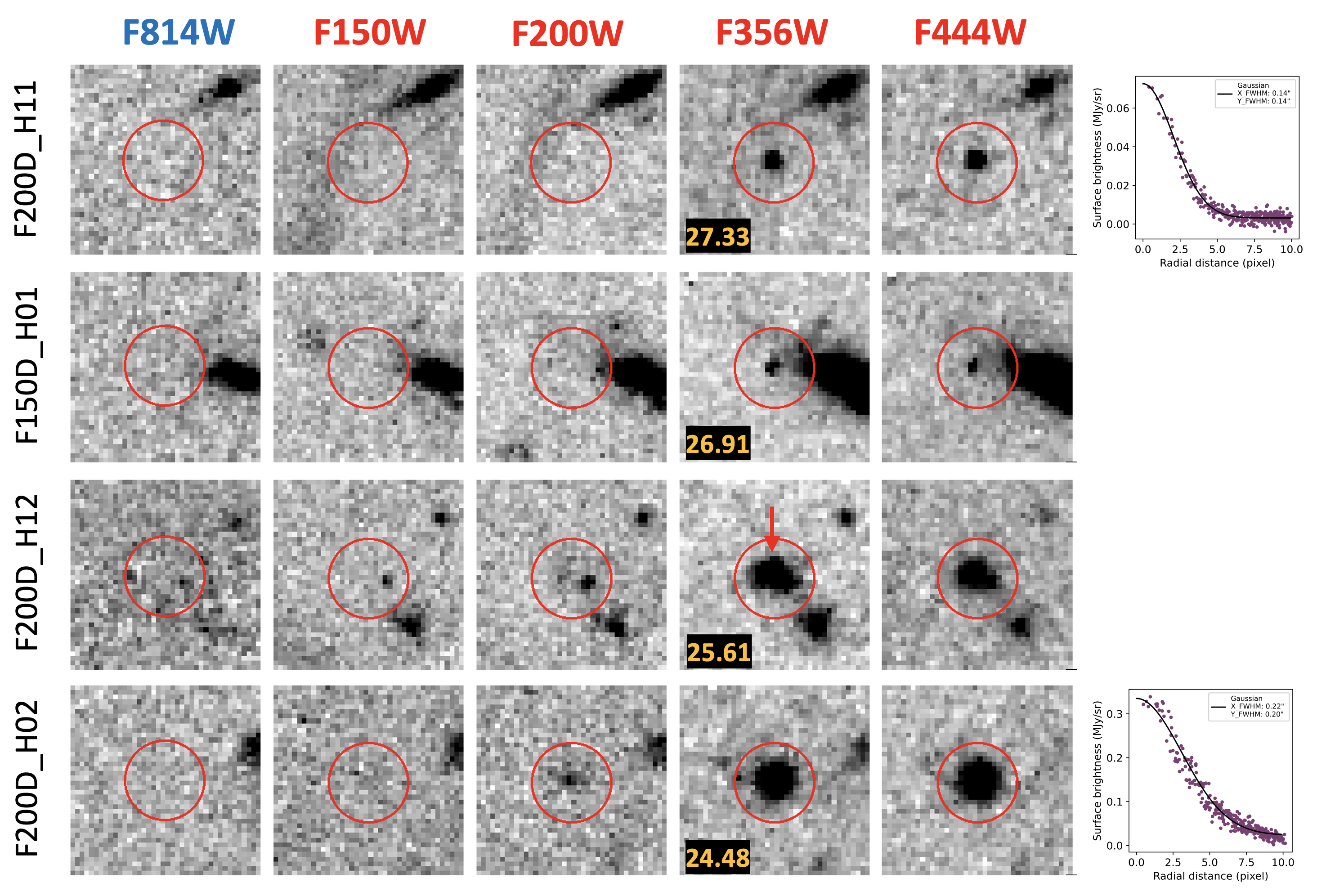}
  \caption{Similar to Figure~\ref{fig:stamps_150d} but for four of the eight
  F200W dropouts. The other four are shown in Figure~\ref{fig:special}.
  The numbers shown on the F356W images are their magnitudes in this band
  ($m_{356}$). \texttt{F200D\_H11} is consistent with being a point source.
}
\label{fig:stamps_200d}
\end{figure*}


\subsection {F200W dropouts}

  There are eight F200W dropouts. The lower-right panel of 
Figure~\ref{fig:2color} shows their colors. The F200W dropout colors are far
from the region occupied by brown dwarfs as well. What makes diagnostics 
difficult with our limited number of passbands is that the color track of 
mid-$z$ old galaxies are between the tracks of the young and old models at high
redshift. Nevertheless, at least one F200W dropout (\texttt{F200D\_H06}) is too
blue in $m_{356}-m_{444}$ to be a mid-$z$ contaminant. 

   Four of the F200W dropouts are shown in Figure~\ref{fig:stamps_200d}. 
The other four are special and are discussed in Section~\ref{s:special}.
Among the four shown in Figure~\ref{fig:stamps_200d}, two are compact
(\texttt{F200D\_H11} and \texttt{H02}). Their 30mas F356W images are fitted
using 2-D Gaussian profiles, which are also shown. One of them, 
\texttt{F200D\_H11}, has FWHM
of 0\farcs14 in both dimensions and is consistent with being a point source.

\begin{figure*}[t]
  \centering
  \includegraphics[height=7in]{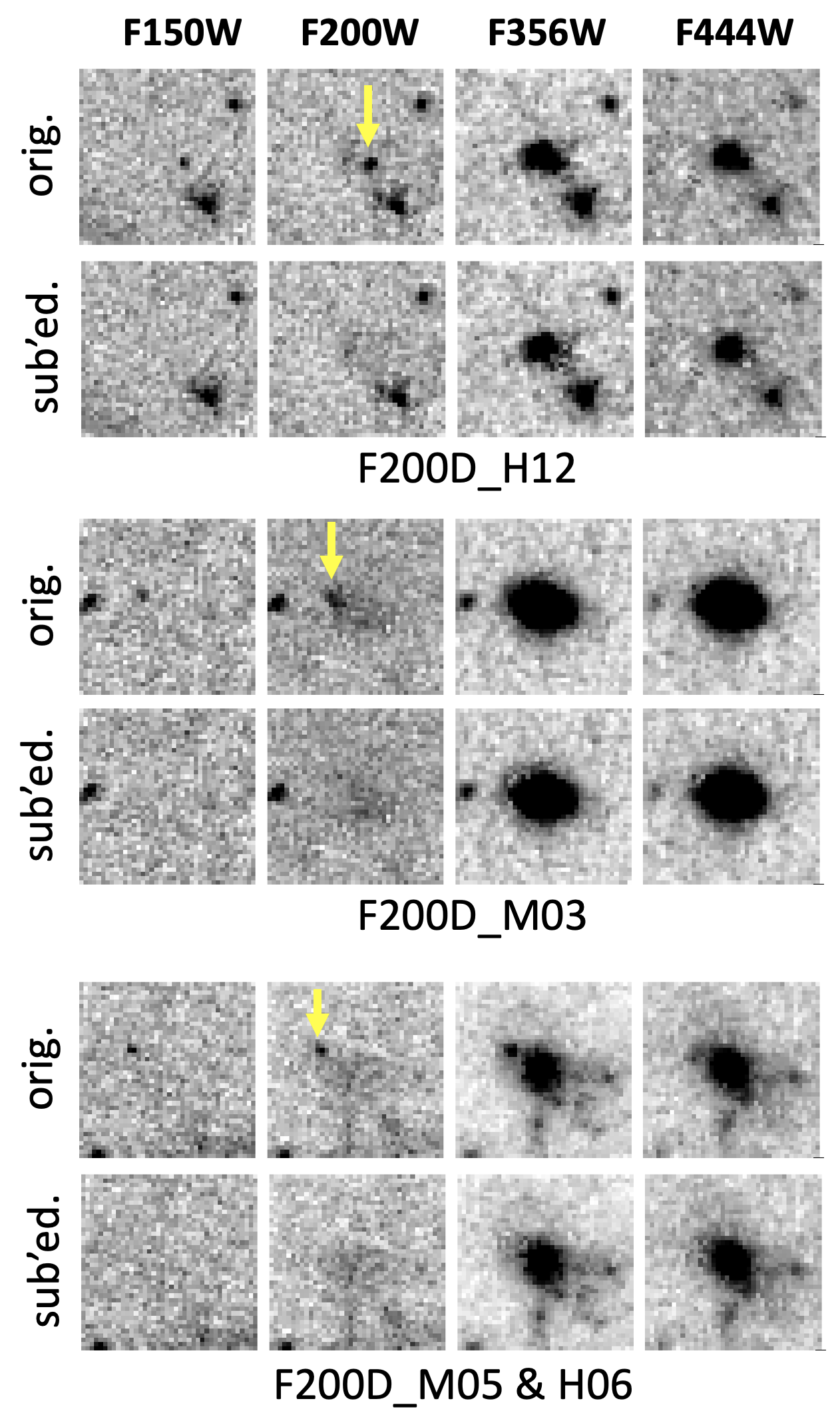}
  \caption{Subtraction of close neighbors for three F200W dropouts that have
unrelated companions. Each object's short ID is below that objects' images, and the passbands are labeled at top.
Stamps are 2\farcs4$\times$2\farcs4 in size with 60~mas pixels.
For each object, the top row shows the original NIRCam negative images, and the bottom
row shows images that have the neighbor (indicated by the yellow 
arrow in the F200W image) subtracted.
}
\label{fig:neighbor_removal}
\end{figure*}

    \texttt{F200D\_H12}, \texttt{M03}, and \texttt{M05} have close
neighbors.  Our photometry was done after  subtracting the neighbors, as shown in Figure~\ref{fig:neighbor_removal}. 

\subsection {Dropouts of Special Interest}
\label{s:special}

\begin{figure*}[t]
  \centering
  \includegraphics[scale=0.35]{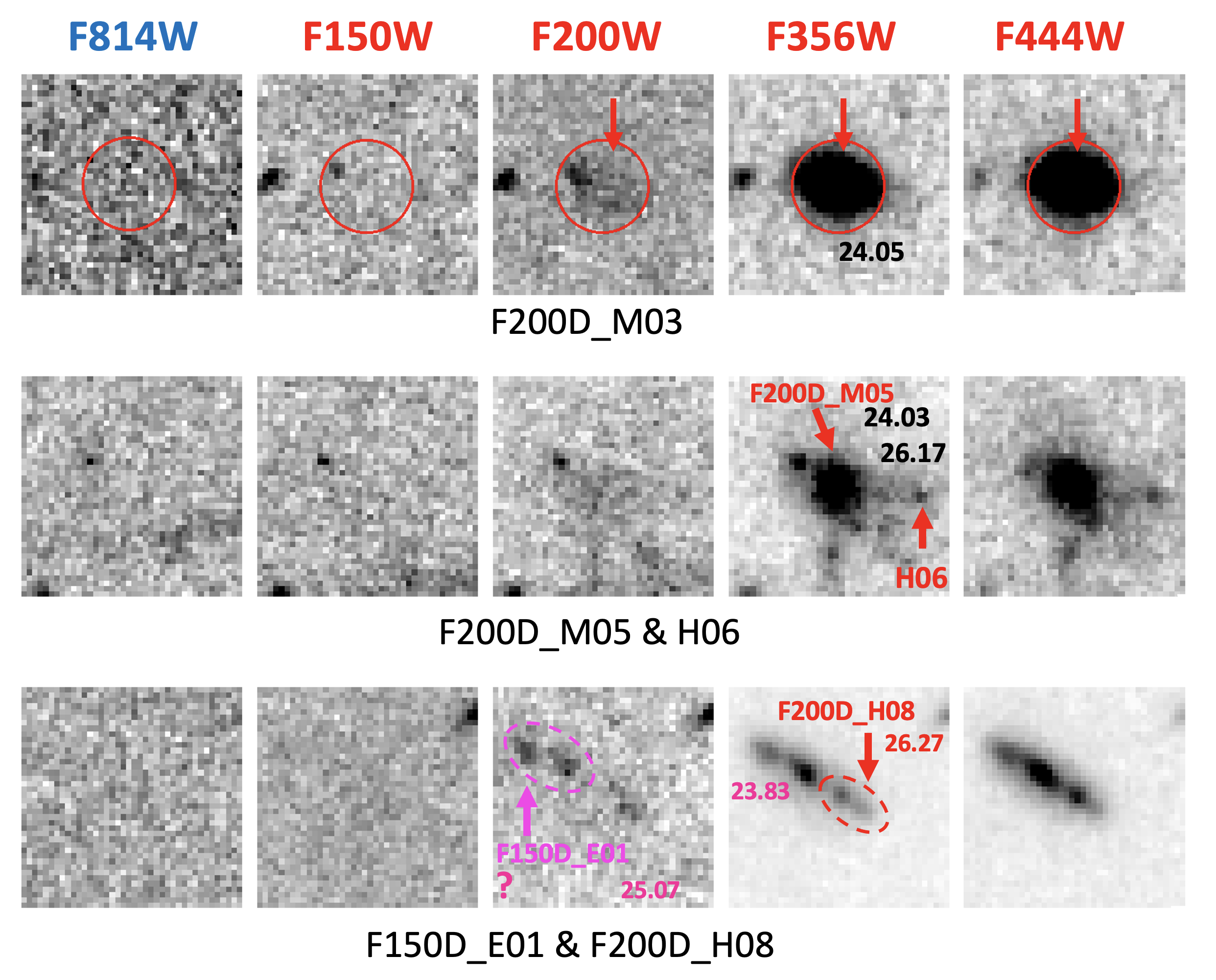}
\caption{Image stamps (2\arcsec.4$\times$2\arcsec.4 in size,
60~mas pixels) of five peculiar 
dropouts. Four of them are F200W dropouts while one is an F150W dropout. Their
IDs are labeled. The labeled magnitudes in F200W or F356W are those in the 
corresponding bands. The one in the top row is the second-brightest in the 
entire sample, and has a close neighbor that is at a low redshift (see also 
Figure~\ref{fig:neighbor_removal}). The middle row shows two F200W dropouts that
are very close neighbors but are very different in brightness and color. The
bottom row shows a bright system that is made of an F150W dropout and an F200W
dropout. 
}
\label{fig:special}
\end{figure*}

    Four F200W dropouts and one F150W dropout are so peculiar that they are 
discussed here separately. Figure~\ref{fig:special} shows their images. 

{   \enumerate
    \item \texttt{F200D\_JWIDF\_M03}\,\, This object was originally selected
as an F150W dropout. Visual inspection shows that it is invisible
in F150W and is only barely detected in F200W and that it has a close neighbor.
After subtracting off this neighbor (see
Figure~\ref{fig:neighbor_removal}), its photometry is consistent with being an
F200W dropout.
It reaches $m_{356}=24.05$ and $m_{444}=23.25$ and is the second-brightest 
object in our entire dropout sample. Its location in color space 
(Figure \ref{fig:2color}) is consistent with either the mid-$z$ old-galaxy
track at $z\approx 8$ or the dusty maximally old track at $z\approx 11$.

    \item \texttt{F200D\_JWIDF\_M05} \& \texttt{H06}\,\, These two objects
are separated by only 0\farcs88 but differ in brightness and color. 
\texttt{H06} is bright, with $m_{356}=26.17$ and $m_{444}=26.31$. \texttt{M05} 
is even brighter, with $m_{356}=24.03$ and $m_{444}=23.95$. 
Similar to 
\texttt{F200D\_JWIDF\_M03}, \texttt{M05} was originally selected as an F150W
dropout, and it also has a close neighbor. After the subtraction of
the neighbor (see Figure~\ref{fig:neighbor_removal}),
it qualifies as an F200W dropout. The location of
\texttt{H06} in  color space is  closest to the track of the young model
at high-$z$, while that of \texttt{M05} is in the ambiguous region where it
can be consistent with either high-$z$ or mid-$z$.

    \item \texttt{F150D\_JWIDF\_E01} \& \texttt{F200D\_JWIDF\_H08}\,\, These
two sources are the strangest because one is an F150W dropout and the other is
an F200W dropout, and yet they are very close to each other. In the F356W
image, this is a system of four blended objects; our source extraction 
identifies the upper two objects as a single source, which is 
\texttt{F150D\_JWIDF\_E01}, and the lower two objects as another single source,
which is \texttt{F200D\_JWIDF\_H08}. The former is the brightest among the 
entire dropout sample but is close to the mid-$z$ old-galaxy contamination 
region in both diagnostic color--color diagrams. If it is indeed due to this 
kind of contamination, it is most likely at $z\approx 3$. The latter source, 
however, is not compatible with such a redshift, which makes their apparent 
association puzzling. This system is similar to the ``chain of five'' system
reported by \citet[][]{Yan2022b}.

}


\section{Discussion}

   The NIRCam F150W and F200W dropouts presented here were selected
by applying the conventional dropout method to $z>11$. In other words, our
result is to say that the JWIDF contains bright $z>11$ galaxy {\it candidates}
similar to those recently reported in other fields.
The new data make the plethora of such objects a more acute problem because
the JWIDF includes even more bright candidates. The faintest F150W dropout has
$m_{200}=27.24$, and the faintest F200W dropout has $m_{356}=27.33$. These
magnitudes are already bright if the sources are indeed at $z>11$. For the sake
of simplicity, let us use 26.5~mag as the fiducial threshold of being 
``very bright'' in this discussion. Five of our F150W dropouts have
$m_{200}<26.5$, and five of our F200W dropouts have $m_{356}<26.5$. This 
implies a cumulative surface density of $\sim$0.7~arcmin$^{-2}$ at 26.5~mag.
Only one dropout (\texttt{F150D\_JWIDF\_H13} with $m_{200}=25.53$) could 
potentially be magnified by gravitational lensing. The brightest four dropouts
have $m_{356}\la24.0$. For reference, $m_{200}$ ($m_{356}$) of 24.0 corresponds
to $M_{\rm UV}=-23.7$ ($-24.12$) at $z=11$ ($z=15$), which is in the luminosity
range of quasars. Two of the dropouts
(\texttt{F150D\_H14} and \texttt{F200D\_H11}) are point-like, which is indeed
consistent with quasar morphology. However, if they are quasars, the inferred 
number density would be orders of magnitude higher than the quasar number
density at $z\approx 7$ \citep[][]{Mortlock2011, Banados2018, Wang2018b, 
Matsuoka2019, Yang2019, Yang2020, Wang2021a}. Furthermore, most of these 
bright objects are extended objects and thus cannot be quasars.

   It is questionable whether known contaminants can explain these dropouts. As
discussed in Section~3, the color diagnostics for the F200W are ambiguous due
to the limited passbands. Diagnostics for the F150W dropouts are based on two
different projections of the color space and are better constrained but still
not conclusive. The results suggest that four F150W dropouts
(\texttt{F150D\_H14}, \texttt{H17}, \texttt{H07}, and \texttt{E01}) have colors
that are also consistent with being contaminants of old galaxies at 
$z\approx 2.6$--8. Interestingly, these four objects (and only these four) also 
satisfy the usual color criterion (approximately $m_{150}-m_{444}\geq 2.3$) for
the ``$H$-band dropout'' galaxies in the literature. (See Section~3.1.) 
Our field also has deep ($\sim$1~$\mu$Jy rms) Jansky Very Large Array 3~GHz 
data (Gim et al., in prep.), which provide some indirect diagnostics. Among all
our 16 dropouts, only two (\texttt{F150D\_H17} and \texttt{E01}) are
detected. They have $S_{\rm 3~GHz}=24.4\pm 3.7$ and $15.2\pm 2.3$~$\mu$Jy, 
respectively, typical of $z\approx 3$ star-forming galaxies, and both are among
the aforementioned possible contaminants. In short, the conservative estimate
is that four of our eight F150W dropouts could be known mid-$z$
contaminants. However, these account for only three of the five that have 
$m_{200}<26.5$. If we attribute the brightness of \texttt{F150D\_H13} to
gravitational lensing, we are still left with \texttt{F150D\_H12} to explain.

   One might argue that some or even all of the dropouts could have been 
``vetoed'' had a bluer NIRCam band been observed.
To test this hypothesis, we used a different PEARLS field that has
eight NIRCam bands to mimic the F150W and F200W dropout selection in the
IDF. We first used only the same four NIRCam bands and an ACS band, and applied
the same selection criteria to select the dropouts. We then added the other
four NIRCam bands to see how many dropouts thus selected would survive. The
details are given in Appendix B. Based on this test, 40\% and 67\% of the
F150W and F200W dropouts in the IDF would survive, respectively.

   One might also argue that there could be some new kinds
of contaminants that we do not consider. Recently, \citet[][]{Zavala2022} 
presented a case where a $z < 6$ dusty starburst mimics the color of an F200W 
dropout. However, such a dropout-like color is mainly due to its old stellar 
population ($\sim$700~Myr old as these authors derived) rather than its being
dusty or star forming. Our color diagnostics have already considered such
mid-$z$, old-age contaminators. Objects with strong nebular emission lines
\citep[e.g.,][]{Wilkins2022} 
might also be suggested as possible contaminators
\citep[e.g.,][]{Naidu2022b}.
However, such objects cannot create dropout-like colors in our color space.
For example, no strong emission lines can conspire to land in 
F200W, F356W, and F444W at the same time to mimic an F150W dropout.
This is also demonstrated in Appendix~B, which shows the SED-fitting 
results of the surviving ``mimicked'' dropouts using the eight-band NIRCam
data. Two different SED fitting tools were utilized, one of which uses a set of
templates including nebular emission lines. The surviving F150W dropouts have
preferred solutions at high-$z$, and at least half of the surviving F200W 
dropouts also have preferred solutions at high-$z$.

   In summary, the very bright F150W and F200W dropouts pose a problem that we
must solve to advance high-$z$ studies with JWST\null. Either these objects 
are due to previously unknown contaminators at $z<11$, or our existing picture
of early galaxy formation needs to be revised. The goal of this work is to
present these troubling and yet interesting objects. Lacking further data, we
are not able to provide more definite interpretations at this time. As the
JWIDF will have two more epochs of NIRCam observations in Cycle~1, 
variability study using the multiple-epoch data might offer some clues. 
However, the most definitive answer will be from spectroscopy. Given their 
brightness, these objects are ideal targets for JWST NIRSpec, and 
obtaining such observations is imperative.

\bigskip

\clearpage

The NIRCam data presented in this paper can be accessed via 
\dataset[10.17909/dh0r-qf34]{https://doi.org/10.17909/dh0r-qf34}
after the proprietary period.

\begin{acknowledgements}
This project is based on observations made with the NASA/ESA/CSA James
Webb Space Telescope and obtained from the Mikulski Archive for Space
Telescopes, which is a collaboration between the Space Telescope Science
Institute (STScI/NASA), the Space Telescope European Coordinating Facility
(ST-ECF/ESA), and the Canadian Astronomy Data Centre (CADC/NRC/CSA).
We thank our Program Coordinator, Tony Roman, for his expert help scheduling
this complex program.

HY and CL acknowledges the partial support from the University of Missouri 
Research Council Grant URC-21-005. SHC, RAW and RAJ acknowledge support from NASA
JWST Interdisciplinary Scientist grants NAG5-12460, NNX14AN10G and 80NSSC18K0200
from GSFC. ZM is supported in-part by the National Science Foundation, grant 
\#1636621. CC is supported by the National Natural Science Foundation of China, 
No. 11803044, 12173045.
CNAW acknowledges funding from the JWST/NIRCam contract NASS-0215 to the
University of Arizona. JFB acknowledges support from NSF Grant No.\ PHY-2012955.
AZ acknowledges support by Grant No. 2020750 from the United States-Israel 
Binational Science Foundation (BSF) and Grant No. 2109066 from the United 
States National Science Foundation (NSF), and by the Ministry of Science 
\& Technology, Israel.

\end{acknowledgements}

\appendix 

\section{Dropout catalog}

   Table ~\ref{tbl:full_dropout_cat} presents the full list of dropouts 
selected in this work. The last four F150W dropouts have colors consistent with being old galaxies at $3\la z \la8$.
   
\begin{table}
\small
\caption{\label{tbl:full_dropout_cat} Catalog of F150W and F200W dropouts in JWIDF
}
\begin{tabular}{ccccccc}
\tableline
ID & R.A. & Decl. & $m_{150}$ & $m_{200}$ & $m_{356}$ & $m_{444}$ \\
\tableline
F150D\_JWIDF\_H11 & 265.007627 & 68.982685 & $>$28.61 & 27.24$\pm$0.14 & 28.09$\pm$0.10 & 28.32$\pm$0.11 \\
F150D\_JWIDF\_H19 & 265.094238 & 68.994877 & 28.19$\pm$0.34 & 26.79$\pm$0.09 & 28.22$\pm$0.10 & 28.85$\pm$0.17 \\
F150D\_JWIDF\_H12 & 264.943505 & 68.982637 & 27.47$\pm$0.48 & 26.15$\pm$0.13 & 25.84$\pm$0.03 & 25.80$\pm$0.03 \\
F150D\_JWIDF\_H13 & 264.943329 & 68.983269 & 26.86$\pm$0.13 & 25.53$\pm$0.04 & 24.99$\pm$0.01 & 24.83$\pm$0.01 \\
\hline
F150D\_JWIDF\_H14 & 265.007586 & 68.983668 & 27.92$\pm$0.44 & 26.70$\pm$0.13 & 25.74$\pm$0.02 & 25.63$\pm$0.01 \\
F150D\_JWIDF\_H17 & 265.008881 & 68.990580 & 26.86$\pm$0.48 & 25.64$\pm$0.14 & 24.51$\pm$0.02 & 24.01$\pm$0.01 \\
F150D\_JWIDF\_H07 & 265.068612 & 68.975125 & 25.52$\pm$0.13 & 24.21$\pm$0.04 & 23.29$\pm$0.00 & 23.04$\pm$0.00 \\
F150D\_JWIDF\_E01 & 264.958298 & 68.985423 & 26.28$\pm$0.29 & 25.07$\pm$0.08 & 23.83$\pm$0.01 & 23.27$\pm$0.01 \\
\tableline
F200D\_JWIDF\_H11 & 265.078543 & 68.994657 & $>$28.71 & $>$28.81 & 27.33$\pm$0.04 & 26.89$\pm$0.03 \\
F200D\_JWIDF\_H01 & 265.040813 & 68.956838 & $>$27.76 & $>$27.89 & 26.91$\pm$0.09 & 26.83$\pm$0.08 \\
F200D\_JWIDF\_H12 & 265.087790 & 68.997920 & $>$27.63 & 27.07$\pm$0.29 & 25.60$\pm$0.02 & 25.52$\pm$0.02 \\
F200D\_JWIDF\_H02 & 265.045779 & 68.957299 & $>$26.87 & 26.52$\pm$0.34 & 24.48$\pm$0.02 & 24.07$\pm$0.01 \\
F200D\_JWIDF\_M05 & 265.146314 & 68.967207 & $>$26.61 & 25.58$\pm$0.19 & 24.03$\pm$0.01 & 23.95$\pm$0.01 \\
F200D\_JWIDF\_H06 & 265.145650 & 68.967187 & $>$27.45 & $>$27.55 & 26.17$\pm$0.05 & 26.31$\pm$0.05 \\
F200D\_JWIDF\_M03 & 264.982466 & 68.962893 & $>$27.45 & 26.55$\pm$0.21 & 24.05$\pm$0.01 & 23.26$\pm$0.00 \\
F200D\_JWIDF\_H08 & 264.957925 & 68.985341 & $>$28.61 & 28.23$\pm$0.35 & 26.27$\pm$0.02 & 25.53$\pm$0.01 \\
\tableline
\end{tabular}
\tablecomments{The coordinates are given in units of degrees and are for J2000. The magnitudes
are SExtractor \texttt{MAG\_ISO} magnitudes, and the limits are 2$\sigma$ 
limits as measured in the \texttt{MAG\_ISO} aperture as defined in the F356W
image. 
}
\end{table}

\newpage

\section{Contamination due to limited NIRCam bands}

  As we only have four NIRCam bands and one (less sensitive) ACS band in this 
field, the interpretation of our F150W and F200W dropouts is more difficult
than for other studies mentioned in Section~1. To better understand our sample,
we used another PEARLS ``blank'' field, the ``NEP Time-domain Field''
(hereafter the ``TDF''), which has eight NIRCam bands reaching comparable 
depths as the IDF \citep[][]{Windhorst2022}. In addition to F150W, F200W,
F356W, and F444W that the IDF has, the TDF also has F090W, F115W, F277W, and
F410M\null. These NIRCam data in the first epoch of the TDF 
($\sim$16~arcmin$^2$ in size) are public. The TDF does not have the ACS F814W
band as the IDF does, but it has the ACS F435W and F606W bands. We therefore
used F606W (deeper of the two) as the optical veto band in the dropout 
selection.

   To test the IDF dropout selection process, we mimicked it in the TDF, using
the same four bands as in the IDF and applying the same color criteria, 
including use of F606W as a veto in the visible wavelength, to select F150W and
F200W dropouts. Then we examined the other four NIRCam bands to check how many
``IDF-mimicked'' dropouts would be rejected. Details of the TDF dropouts will
be given in a future paper, and here we only present the results relevant to
this test.

   The initial selection produced 
five F150W dropouts, and three of them were rejected by the formal $S/N \geq 2$
detections in F115W and/or F090W or by possible weak detections judged by eye. 
Similarly, there were nine F200W dropouts selected initially, and three were 
rejected after using F115W and F090W\null. This implies that the contamination
rates due to limited NIRCam bands are 60\% and 33\% for the initial F150W and
F200W dropouts, respectively. Image stamps of all 14 objects are shown in
Figures~\ref{fig:TmI_stamps_f150d} and~\ref{fig:TmI_stamps_f200d}.

   To further study whether the surviving dropouts are consistent with being at 
high-$z$, we fitted their SEDs to derive their photometric redshifts
($z_{\rm ph}$). However, we caution that $z_{\rm ph}>11$ for a 
particular dropout should not be taken as a {\em confirmation} of its being at 
high-$z$. This is because of the statistical nature of $z_{\rm ph}$. 
``Catastrophic failures'' of $z_{\rm ph}$ are always possible. By the same 
token, $z_{\rm ph}<11$ for a particular dropout should not necessarily exclude
it from the sample. Dropout selection and $z_{\rm ph}$ selection of high-$z$
candidates are two different methods, and one is not superior to the other. 
Nevertheless, the likelihood of a candidate being at high-$z$ increases if it
passes the selection of both methods.

   As SED fitting depends on both the adopted templates and the method
used, we took two different approaches. One was using \texttt{Le Phare}
\citep[][]{Arnouts1999,Ilbert2006} to fit our dropouts to galaxy templates
based on the BC03 models. The templates were constructed assuming exponentially
declining star formation histories in the form of SFR $\propto e^{-t/\tau}$, 
where $\tau$ ranged from 0 to 13~Gyr (0 for SSP and 13~Gyr to approximate 
constant star formation). These models use the Chabrier initial mass function
\cite[][]{Chabrier2003}. We adopted the Calzetti extinction law, with $E(B-V)$
ranging from 0 to 1.0 mag. 
The other approach was using \texttt{EAZY-py},
\footnote{\url{https://github.com/gbrammer/eazy-py}} which is the latest
implementation of \texttt{EAZY} \cite[][]{Brammer2008}. We adopted the
``FSPS 12'' template set, which include nebular emission. We  
modified the code to use flux density upper limits. 

   The results are shown in Figure \ref{fig:TmI_sedfitting}.
For the two surviving F150W dropouts,
both methods give the best solutions at $z_{\rm ph}>10$: one has
best-fit $z_{\rm ph}>13$, while the other has best-fit $z_{\rm ph}=10.4$ and 
11.9 by \texttt{Le Phare} and \texttt{EAZY}, respectively. The probability
distribution functions (PDFs) are all narrow and prefer high-$z$ solutions.
For the six surviving F200W dropouts, both methods give  best-fit 
$z_{\rm ph}\gtrsim 20$ for three objects and $z_{\rm ph}\lesssim 9$ for the
other three. In general, the PDFs are rather broad and span both the high-$z$
and low-$z$ ranges, especially in the \texttt{EAZY} results. A conservative
statement is that neither method can rule out high-$z$ solutions for at least 
three of the six surviving F200W dropouts.

    Through the aforementioned assessment, we conclude that {\it at least} a 
significant fraction ($>$33\%) of the F150W and F200W dropouts selected in the
IDF are legitimate high-$z$ candidates. The other dropouts might not be
at high-$z$ but should be further studied to understand
the contaminating population(s) for high-$z$ selection in the JWST era. 

\setcounter{figure}{0}
\renewcommand{\thefigure}{B\arabic{figure}}

\begin{figure*}[t]
  \centering
  \includegraphics[width=\textwidth]{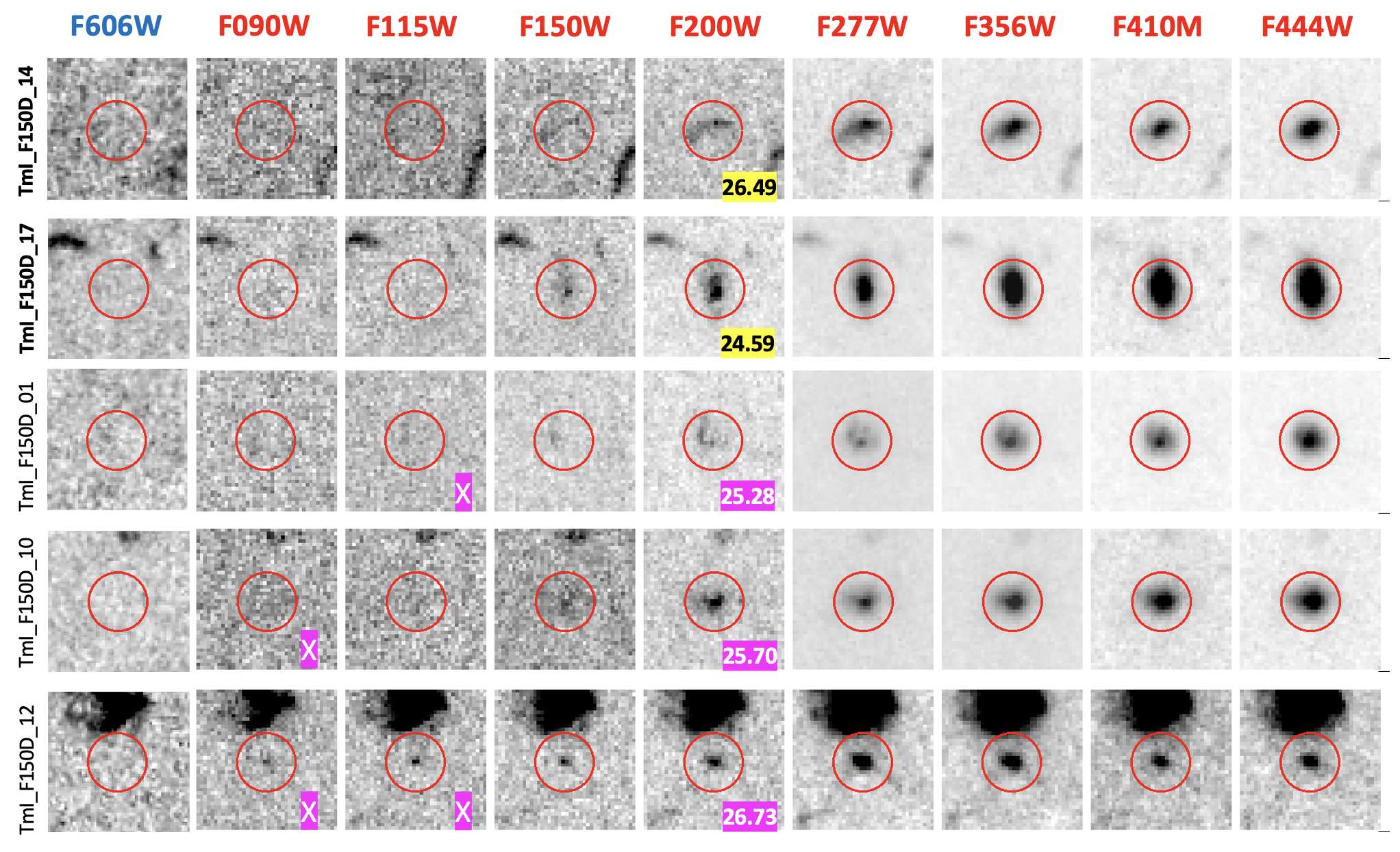}
\caption{Nine-band image stamps of the five F150W dropout candidates selected
    in the TDF\null. The images are 2\farcs4$\times$2\farcs4 in size and are
    oriented north-up and east-left. The candidates are centered on the images
    and are indicated by red circles with radii 0\farcs5. These F150W 
    candidates were selected using only the same four NIRCam bands as in the
    IDF plus the HST ACS F606W band.
    (Short IDs of sources are ``TmI'' for ``TDF mimicking IDF.'') 
    The numbers shown on the F200W images are their magnitudes in this band.
    The two dropouts that survive after incorporating the other four NIRCam 
    bands (especially F090W and F115W as the veto bands) are in the top two 
    rows. The bottom three rows show sources rejected by their formal 
    $S/N \geq 2$ detections in F115W and/or F090W or by the visual inspection
    in these two bands (marked by ``X'').
}
\label{fig:TmI_stamps_f150d}
\end{figure*}

\begin{figure*}[t]
  \centering
  \includegraphics[width=\textwidth]{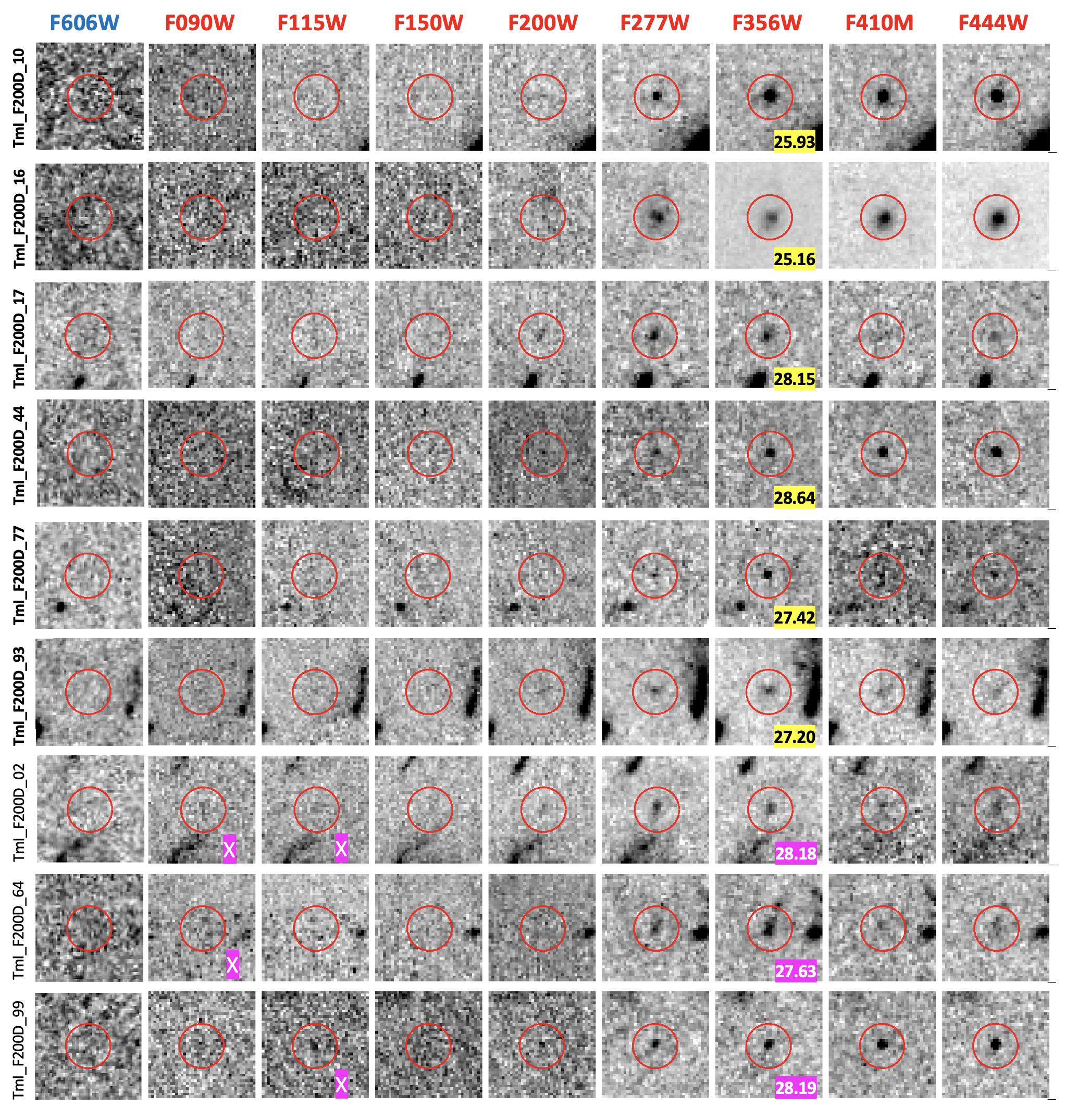}
\caption{Similar to Figure~\ref{fig:TmI_stamps_f150d} but for the nine
TmI F200W candidates. The numbers shown on the F356W images are their magnitudes
in this band. After incorporating the other four NIRCam bands, the objects in
the top six rows survive, and those in the bottom three rows are rejected by 
images marked with ``X.''
}
\label{fig:TmI_stamps_f200d}
\end{figure*}

\begin{figure*}[t]
  \centering
  \includegraphics[width=0.8\textwidth]{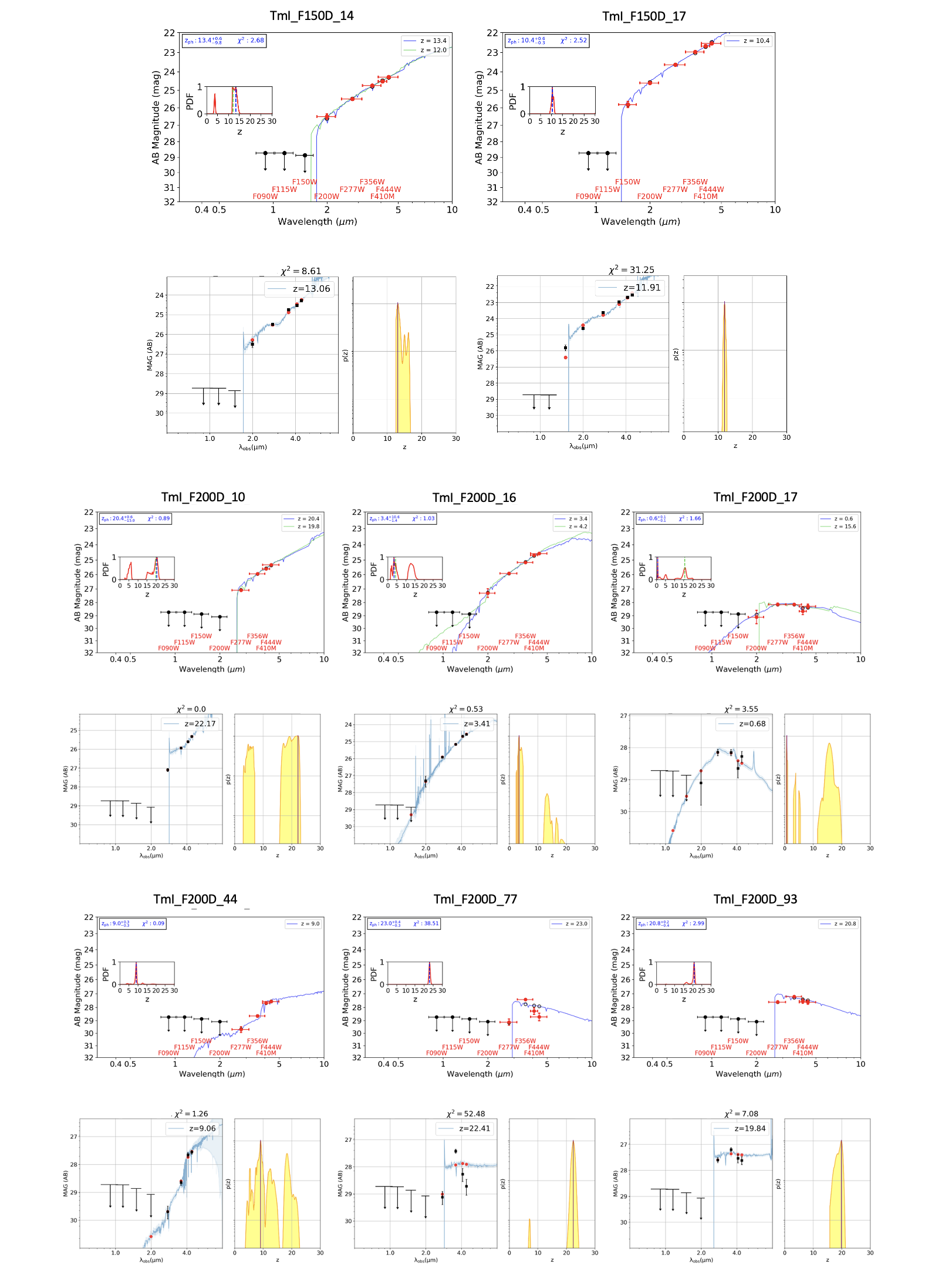}
\caption{SED fitting results of the surviving TmI F150W and F200W dropouts. The
first two rows show the two F150W dropouts 
with the \texttt{Le Phare} results above and the 
\texttt{EAZY} results below. Source names are above each \texttt{Le Phare} panel.  The remaining rows show
the six F200W dropouts in the same arrangement.
The derived photometric redshifts and $\chi^2$ values  are given near the top of each panel. The $\chi^2$ values are the
raw values, i.e., not the reduced $\chi^2$.
In each \texttt{Le Phare} panel, red circles and black upper limits show the data. Blue curves are the  best-fit models corresponding to the first peak
of the redshift PDFs. The PDFs themselves are shown as insets.
The green curves, when present, are the best-fit model 
corresponding to the second peak of the PDF\null. In the panels showing the
\texttt{EAZY} results, the black symbols show the data, and the curves
show the best-fit models. The red circles mark the synthesized magnitudes based
on the best-fit models. The PDFs are shown (with yellow fill) next to the model fits.
}
\label{fig:TmI_sedfitting}
\end{figure*}


\bibliographystyle{aasjournal.bst}


\end{document}